\documentclass[prl,showpacs, twocolumn, aps,superscriptaddress,floatfix]{revtex4}
\usepackage{amsmath}
\usepackage{amssymb}
\usepackage{bm}
\usepackage{graphicx}
\usepackage{color}
\usepackage[svgnames]{xcolor}
\usepackage{soul}
\usepackage{cancel}
\usepackage{hyperref}
\usepackage{verbatim}

\begin{document}

\title{Can bilayer graphene become a fractional metal?}

\author{A.O. Sboychakov}
\affiliation{Institute for Theoretical and Applied Electrodynamics, Russian
Academy of Sciences, 125412 Moscow, Russia}

\author{A.L. Rakhmanov}
\affiliation{Institute for Theoretical and Applied Electrodynamics, Russian
Academy of Sciences, 125412 Moscow, Russia}

\author{A.V. Rozhkov}
\affiliation{Institute for Theoretical and Applied Electrodynamics, Russian
Academy of Sciences, 125412 Moscow, Russia}

\author{Franco Nori}
\affiliation{Advanced Science Institute, RIKEN, Wako-shi, Saitama,
351-0198, Japan}
\affiliation{Department of Physics, University of Michigan, Ann
Arbor, MI 48109-1040, USA}

\begin{abstract}
It is known that electron interactions can cause a perfect spin polarization of the Fermi surface of a metal. In such a situation only half of the non-interacting Fermi surface is available, and thus this phase is commonly referred to as a `half-metal'. Here we argue that, in multi-band electronic systems with nesting, further `fractionalization' of the Fermi surface is possible. Taking the AA~bilayer graphene as a convenient test case, we demonstrate that, under suitable conditions imposed on the electron interactions, doped AA~bilayer graphene can host a `quarter-metal' state. In such a state, only one quarter of the non-interacting Fermi surface (Fermi contour) reaches the Fermi energy. At higher doping level, other `fractional' metals can emerge. We briefly analyze the transport properties of these proposed phases.
\end{abstract}


\date{\today}

\maketitle

\textit{Introduction}.--- In usual metals, the total spin polarization
of the charge carriers at the Fermi surface is zero. A strong electron-electron
interaction can lift the spin degeneracy, and induce spin polarization of the
states at the Fermi surface. In the extreme case of the so-called
half-metals~\cite{first_half_met1983,half_met_review2008,hu2012half}, this
polarization is perfect: all states at the Fermi energy have identical spin
projection. Indeed, various rather different systems with transition-metal atoms are found to be half-metals~\cite{nimnsb_exp1990,lasrmno_half_met_exp1998,cro2_half_met_exp2001,co2mnsi_half_met_exp2014}. The existence of spin-polarized currents in these half-metals makes them promising materials for applications in spintronics~\cite{review_spintronics2004,hu2012half}. Several papers~\cite{metal_free_hm2012,meta_free_hm2014,son2006half,kan2012half,huang2010intrinsic} predict half-metallicity in carbon-based systems. The half-metals free of heavy atoms could be of interest for bio-compatible applications and carbon-based electronics~\cite{soriano2010,plastic_electr2010,Avouris2007,meso_review,chinese_phys_silicene2014,bilayer_review2016}.

In previous works~\cite{our_hmet_prl2017,our_hmet_prb2018} we have proposed a mechanism for half-metallicity in electronic systems with weak interactions. This requires the existence of two Fermi surface sheets with nesting between them. These sheets are referred below as having `electron/hole charge flavors'~\cite{note_on_valley}. When doped, the spin-density wave (SDW) or charge-density wave (CDW) insulator state in such a model is replaced by this type of  half-metallic state.

In a multi-band system with nesting, besides spin, an additional discrete quantum number $\xi$  emerges, enumerating pairs of nested Fermi surface sheets. In such a situation, one may wonder if a many-body state with an additional polarization with respect to $\xi$ could be realized. The stability of this peculiar conducting state, which we call below ``fractional metal'' (FraM), is the main topic of this paper.

It follows from the FraM definition that only a material with sufficiently
complex multi-sheet Fermi surface with nesting might host a FraM phase. This
requirement makes AA bilayer graphene (AA-BLG) a promising candidate to be
a FraM. The AA-BLG is less studied than the Bernal stacked (AB) bilayer.
Yet, AA-BLG samples have been manufactured~\cite{aa_graphit_roy1998,aa_graphite_lee2008,liu_aa_exp2009,
borysiuk2011_aa}. Moreover, progress in van~der~Waals heterostructures
fabrication~\cite{geim_vdW_heter2013} allows one to hope that more
efforts will be undertaken in the direction of producing
high-quality AA-BLG samples. As in other graphene structures, the
low-energy states of the AA-BLG can be classified by their proximity to
either the ${\bf K}_1$ or ${\bf K}_2$ Dirac point. A given Dirac point is
encircled by an electron Fermi surface sheet and a hole sheet; altogether
there are four Fermi surface sheets in the whole Brillouin zone. We argue
that, for such a degenerate Fermi surface structure and under rather common
assumptions about the electron-electron coupling, doped AA-BLG could enter
the FraM phase. We investigate the stability of this phase and also briefly
discuss its most immediate properties, such as transport of spin and valley
quanta, and peculiar features of superconductivity.

\textit{Model}.--- The electronic properties of AA-BLG are described by the Hamiltonian
$\hat{H}=\hat{H}_0+\hat{H}_{\textrm{int}}$, where $\hat{H}_0$
is the single-electron part and $\hat{H}_{\textrm{int}}$
corresponds to the interaction between quasiparticles. For AA-BLG~\cite{bilayer_review2016}:
\begin{equation}
\label{SingleParticleHam}
\!\!\hat{H}_0
\! =\!
-t\!\!\!\sum_{\langle \mathbf{mn}\rangle l\sigma}\!\!\!\! 	
	d^\dag_{\mathbf{m}l0\sigma}d^{\phantom{\dag}}_{\mathbf{n}l1\sigma}
\!-\!
t_0\!\sum_{\mathbf{n}a\sigma}
d^\dag_{\mathbf{n}0a\sigma}d^{\phantom{\dag}}_{\mathbf{n}1a\sigma}
+
{\rm H.c.}-\mu n\,.
\end{equation}
Here $d^\dag_{\mathbf{m}la\sigma}$ ($d^{\phantom{\dag}}_{\mathbf{m}la\sigma}$) is the creation (annihilation) operator of an electron with spin projection $\sigma$ in layer $l$ [$l=0$ ($l=1$) corresponds to upper (lower) layer] on sublattice $a$ [$a=0$ ($a=1$) represents sublattice $A$ ($B$)] at the position $\mathbf{m}$. Also, $n=\sum_{\mathbf{n}la\sigma} 	 d^\dag_{\mathbf{m}la\sigma}d^{\phantom{\dag}}_{\mathbf{m}la\sigma}$ is the total charge density, $\mu$ is the chemical potential, and $\langle...\rangle$ denotes nearest-neighbor pairs. The amplitude $t=2.57$\,eV ($t_0=0.36$\,eV) describes the in-plane (inter-plane) nearest-neighbor hopping. $\hat{H}_0$ can be readily diagonalized in a new basis $\gamma^{\phantom{\dag}}_{\mathbf{k}\ell\sigma}$ ($\ell=1, \ldots, 4$):
\begin{equation}
\label{HamDiag}
\hat{H}_0=\sum_{\mathbf{k}\ell\sigma}
	{\left(\varepsilon^{(\ell)}_{0\mathbf{k}}-\mu\right)
		\gamma^\dag_{\mathbf{k}\ell\sigma}
		\gamma^{\phantom{\dag}}_{\mathbf{k}\ell\sigma}},
\end{equation}
where $\mathbf{k}$ is the momentum; the eigenenergies and eigenoperators are
\begin{eqnarray}
\varepsilon^{(1)}_{0\mathbf{k}}\!\!&=&\!\!-t_0-t\zeta_\mathbf{k},\quad
\varepsilon^{(2)}_{0\mathbf{k}}= -t_0+t\zeta_\mathbf{k},\nonumber\\
\varepsilon^{(3)}_{0\mathbf{k}} \!\!&=&\!\! +t_0-t\zeta_\mathbf{k},\quad
\varepsilon^{(4)}_{0\mathbf{k}}=+t_0+t\zeta_\mathbf{k},\quad
\\
d_{\mathbf{k}la\sigma}\!\!&=&\!\!\exp{(-ai\varphi_\mathbf{k})}
\!\left[
	\gamma_{\mathbf{k}1\sigma}
	\!+\!
	(-1)^a\gamma_{\mathbf{k}2\sigma}
	+
\right.
\nonumber
\\
&&
\left.
	(-1)^l\gamma_{\mathbf{k}3\sigma}
	\!+\!	
	(-1)^{a+l}\gamma_{\mathbf{k}4\sigma}\right]/2.
\label{eigen_En_Fn}
\end{eqnarray}
In Eq.~\eqref{eigen_En_Fn}, $\varphi_\mathbf{k}=\textrm{arg}(f_\mathbf{k})$, $\zeta_\mathbf{k}=|f_\mathbf{k}|$, where $f_\mathbf{k} = 1+2\exp{\left({3}ik_xa_0/2\right)} \cos{(\sqrt{3}k_ya_0/2)}$, and $a_0=1.42$~{\AA} is the in-plane carbon-carbon distance. The band $\ell = 2$ (band $\ell = 3$) crosses the Fermi level and forms two electron (two hole) Fermi surface sheets, one centered at the Dirac point $\mathbf{K}_{1}=2\pi(\sqrt{3},1)/3\sqrt{3}a_0$, and another at $\mathbf{K}_{2}=2\pi(\sqrt{3},-1)/3\sqrt{3}a_0$. To distinguish electron and hole Fermi surface sheets, we introduce the charge flavor index $\nu = (-1)^\ell$: it equals $\nu = 1$ ($\nu = - 1$) for electrons (holes). If we label~\cite{note_on_valley} the graphene valley ${\bf K}_1$
(valley ${\bf K}_2$) by $\xi = +1$ (by $\xi = -1$), any sheet can be uniquely identified by values of $\nu$ and $\xi$. Since all sheets are circles of identical radius $k_{F0}=2t_0/3ta_0$, we have two nesting vectors: $\mathbf{0}$ and $\mathbf{Q}_0=\mathbf{K}_1-\mathbf{K}_2$.

The Coulomb interaction between electrons is
\begin{eqnarray}
\label{interaction}
\hat{H}_{\textrm{int}}
=
\frac{1}{2N_c}\!\!\!
\sum_{\mathbf{kk}'\mathbf{q}la\atop l'a'\sigma\sigma'}\!\!\!
	V^{ll'}_{aa'}(\mathbf{q})
	d^\dag_{\mathbf{k}la\sigma}
	d^{\phantom{\dag}}_{\mathbf{k}+\mathbf{q}la\sigma}
	d^\dag_{\mathbf{k}'l'a'\sigma'}
	d^{\phantom{\dag}}_{\mathbf{k}'-\mathbf{q}l'a'\sigma'},
\end{eqnarray}
where
$N_c$
is the number of elementary cells in the sample and
$V^{ll'}_{aa'}(\mathbf{q})$
is the Fourier transform of
\begin{eqnarray}
\label{potential}
V^{ll'}_{aa'}({\bf r})
=
V_C\left(\sqrt{[{\bf r}+ (a-a')\bm{\delta}_1]^2 + (l-l')^2 D^2 }\right).
\end{eqnarray}
Here, $V_C (|{\bf r}|)$ is the screened Coulomb potential, $\bm{\delta}_1=(a_0,\,0)$, and $D=3.3$~{\AA} is the inter-layer distance. The dependence of the interaction on various indices accounts for different distances between electrons at different sublattices and/or layers.

\textit{Mean field approach}.--- Theory predicts~\cite{aa_graph_prl2012,aa_graph_pha_sep_sboycha2013, aa_graph_BreyFertig2013,aa_graph_akzyanov2014} that the electron repulsion converts the electronic ``liquid'' of the AA-BLG into a SDW insulator. The SDW order is characterized by non-zero values of $\langle\gamma^\dag_{\mathbf{k}2\sigma}\gamma^{\phantom{\dag}}_{\mathbf{k}3\bar{\sigma}}\rangle$ and $\langle\gamma^\dag_{\mathbf{k}1\sigma}\gamma^{\phantom{\dag}}_{\mathbf{k}4\bar{\sigma}}\rangle$, which describe excitonic pairs with vanishing total momentum. It is possible to define a different order parameter oscillating in space with the wave vector ${\bf Q}_0$, e.g., $\langle\gamma^\dag_{\mathbf{k+Q}_02\sigma}\gamma^{\phantom{\dag}}_{\mathbf{k}3\bar{\sigma}}\rangle$.
However, the oscillating order parameter has lower coupling constant, because it cannot interact with another oscillating order parameter unless they have opposite momenta. This condition strongly reduces the effective coupling constant. As a result, such a phase has higher energy, and we will not consider it here. Switching to band operators $\gamma$ and neglecting the terms irrelevant
to the mean field approximation, we transform Eq.~\eqref{interaction} and write
\begin{equation}
\hat{H}_{\textrm{int}}= \hat{H}^{(1)} + \hat{H}^{(2)} + \hat{H}^{(3)} + \hat{H}^{(4)},
\end{equation}
where
\begin{eqnarray}
\label{eq::int_gamma_subst}
\hat{H}^{(1)} &=&
	-\frac{1}{N_c}
	\sum_{\mathbf{kp}  \sigma }
		V^{(1)}_{{\bf kp}} \left[
			(\gamma_{\mathbf{k}1\sigma}^\dag
			\gamma_{\mathbf{k}4\bar{\sigma}}^{\phantom{\dag}})
			(\gamma_{\mathbf{p}4\bar{\sigma}}^\dag
			\gamma_{\mathbf{p}1\sigma}^{\phantom{\dag}})
\right.\nonumber\\
&&
\left.
			+
			(\gamma_{\mathbf{k}3\bar{\sigma}}^\dag
			\gamma_{\mathbf{k}2\sigma}^{\phantom{\dag}})
			(\gamma_{\mathbf{p}2\sigma}^\dag
			\gamma_{\mathbf{p}3\bar{\sigma}}^{\phantom{\dag}})
	\right],\\
\label{H2}
\hat{H}^{(2)}&=&
	-\frac{1}{2N_c}
	\sum_{\mathbf{kp}  \sigma }
		V^{(2)}_{{\bf kp}} \left[
			(\gamma_{\mathbf{k}1\sigma}^\dag
			\gamma_{\mathbf{k}4\bar{\sigma}}^{\phantom{\dag}})
			(\gamma_{\mathbf{p}1\bar{\sigma}}^\dag
			\gamma_{\mathbf{p}4\sigma}^{\phantom{\dag}})
\right.\nonumber\\
&&
\left.
			+
			(\gamma_{\mathbf{k}2\sigma}^\dag
			\gamma_{\mathbf{k}3\bar{\sigma}}^{\phantom{\dag}})
			(\gamma_{\mathbf{p}2\bar{\sigma}}^\dag
			\gamma_{\mathbf{p}3\sigma}^{\phantom{\dag}})
			+
			{\rm H.c.}
	\right],\\
\hat{H}^{(3)}&=&
	-\frac{1}{N_c}
	\sum_{\mathbf{kp}  \sigma }
		V^{(3)}_{{\bf kp}} \left[
			(\gamma_{\mathbf{k}1\sigma}^\dag
			\gamma_{\mathbf{k}4\bar{\sigma}}^{\phantom{\dag}})
			(\gamma_{\mathbf{p}3\bar{\sigma}}^\dag
			\gamma_{\mathbf{p}2\sigma}^{\phantom{\dag}})
\right.\nonumber\\
&&
\left.
			+
			(\gamma_{\mathbf{k}2\sigma}^\dag
			\gamma_{\mathbf{k}3\bar{\sigma}}^{\phantom{\dag}})
			(\gamma_{\mathbf{p}4\bar{\sigma}}^\dag
			\gamma_{\mathbf{p}1\sigma}^{\phantom{\dag}})
	\right],\\
\label{H4}
\hat{H}^{(4)}&=&
	-\frac{1}{2N_c}
	\sum_{\mathbf{kp}  \sigma }
		V^{(4)}_{{\bf kp}} \left[
			(\gamma_{\mathbf{k}1\sigma}^\dag
			\gamma_{\mathbf{k}4\bar{\sigma}}^{\phantom{\dag}})
			(\gamma_{\mathbf{p}2\bar{\sigma}}^\dag
			\gamma_{\mathbf{p}3\sigma}^{\phantom{\dag}})
\right.\nonumber\\
&&
\left.
			+
			(\gamma_{\mathbf{k}2\sigma}^\dag
			\gamma_{\mathbf{k}3\bar{\sigma}}^{\phantom{\dag}})
			(\gamma_{\mathbf{p}1\bar{\sigma}}^\dag
			\gamma_{\mathbf{p}4\sigma}^{\phantom{\dag}})
			+
			{\rm H.c.}
		\right],
\end{eqnarray}
with the coupling constants $V^{(1,2,3,4)}_{{\bf kp}}$ defined by
\begin{eqnarray}
\label{V13}
V^{(1,3)}_{{\bf kp}} = \frac{1}{8}
\left[
	V^{00}_{AA}\!+\!V^{10}_{AA}
	\!\pm\!
	\left(V^{00}_{AB}\!+\!V^{10}_{AB}\right)e^{-i\Delta\varphi}
	\!+\!
	{\rm C.c.}
\right]\!,
\quad
\\
\label{V24}
V^{(2,4)}_{{\bf kp}} = \frac{1}{8}
\left[
	V^{00}_{AA}\!-\!V^{10}_{AA}\!
	\mp\!
	\left(V^{00}_{AB}\!-\!V^{10}_{AB}\right)e^{-i\Delta\varphi}
	\!+\!
	{\rm C.c.}
\right]\!.
\quad
\end{eqnarray}
Here $V^{ll'}_{aa'} = V^{ll'}_{aa'}(\mathbf{k}-\mathbf{p})
=V^{ll'}_{a'a}(\mathbf{p}-\mathbf{k})$,
and $\Delta \varphi = \Delta \varphi_{\mathbf{kp}} = \varphi_{\mathbf{k}} - \varphi_{\mathbf{p}}$.
One can assume~\cite{Nandkishore2010,Nandkishore2010b}
that intra-layer and inter-layer interactions in a graphene bilayer are approximately equal (at small momentum): $V^{00}_{aa'} \approx V^{10}_{aa'}$. In such a limit, we have in the first approximation
\begin{eqnarray}
\label{matrix_V13}
V^{(1,3)}_{{\bf kp}}
\approx
\frac{1}{2} V_C(\mathbf{k-p})
	\left[1\pm \cos (\Delta\varphi_\mathbf{kp})\right],
\quad
V^{(2,4)}_{{\bf kp}} \approx 0.
\quad
\end{eqnarray}
Thus, the interaction can be approximated as
$\hat{H}_{\textrm{int}} \approx \hat{H}^{(1)} + \hat{H}^{(3)}$.
We analyze this Hamiltonian using mean field theory, and the terms
$\hat{H}^{(2,4)}$ will be taken into account perturbatively. The mean field version of
$\hat{H}_{\textrm{int}}$
is
\begin{eqnarray}
\label{eq::int_MF_fin}
\hat{H}_{\textrm{int}}^{\rm MF}
\!=
\!-\!
\sum_{\mathbf{p}  \sigma }
		\tilde \Delta_{{\bf p} \sigma}
		\gamma_{\mathbf{p}4\bar{\sigma}}^\dag
		\gamma_{\mathbf{p}1\sigma}^{\phantom{\dag}}
		\!\!+\!
		\Delta_{{\bf p} \sigma}
		\gamma_{\mathbf{p}3\bar{\sigma}}^\dag
		\gamma_{\mathbf{p}2\sigma}^{\phantom{\dag}}
		\!+\!
		{\rm H.c.}
\!+\!  B,
\quad
\end{eqnarray}
where
\begin{eqnarray}
\label{Delta}
\nonumber
\Delta_{\mathbf{k}\sigma}
&\!\!=\!\!&
\frac{1}{N_c}\!
\sum_{\mathbf{p}}
	\!\left[
		V^{(1)*}_\mathbf{pk}
		\langle
			\gamma^\dag_{\mathbf{p}2\sigma}
			\gamma_{\mathbf{p}3\bar{\sigma}}^{\phantom{\dag}}
		\rangle
		\!+\!
  		V^{(3)}_{\mathbf{pk}}
		\langle
			\gamma^\dag_{\mathbf{p}1\sigma}
			\gamma_{\mathbf{p}4\bar{\sigma}}^{\phantom{\dag}}
		\rangle
	\right],
\\
\nonumber
\tilde{\Delta}_{\mathbf{k} \sigma}
&\!\!=\!\!&
\frac{1}{N_c}\!
\sum_{\mathbf{p}}
  	\left[
		V^{(1)}_{\mathbf{pk}}
		\langle
			\gamma^\dag_{\mathbf{p}1\sigma}
			\gamma_{\mathbf{p}4\bar{\sigma}}^{\phantom{\dag}}
		\rangle
		\!+\!
  		V^{(3)}_{\mathbf{pk}}
		\langle
			\gamma^\dag_{\mathbf{p}2\sigma}
			\gamma_{\mathbf{p}3\bar{\sigma}}^{\phantom{\dag}}
		\rangle
	\right],
\\
B
&\!\!=\!\!&
\sum_{\mathbf{p} \sigma }
	\left[
		\Delta_{\mathbf{p} \sigma}
		\langle
			\gamma^\dag_{\mathbf{p}3\bar{\sigma}}
			\gamma_{\mathbf{p}2\sigma}^{\phantom{\dag}}
		\rangle
  		\!+\!
		\tilde{\Delta}_{\mathbf{p} \sigma}
		\langle
			\gamma^\dag_{\mathbf{p}4\bar{\sigma}}
			\gamma_{\mathbf{p}1\sigma}^{\phantom{\dag}}
		\rangle
	\right].\phantom{aaa}
\end{eqnarray}

The spectrum of the mean-field Hamiltonian can be easily derived:
\begin{eqnarray}
\label{spectrum}
\!\!\!\!&&E^{(2,3)}_{{\bf k} \sigma} = \mp E^{\rm m}_{\bf k \sigma },
\qquad
E^{(1,4)}_{{\bf k} \sigma} = \mp E^{\rm h}_{\bf k \sigma },\\
\nonumber
\!\!\!\!&&E^{\rm m}_{\bf k \sigma }\!\!=\!\!\sqrt{ |\Delta_{\bf k \sigma}|^2\! +\! (t_0\! -\! t \zeta_{\bf k})^2 },\,\,
E^{\rm h}_{\bf k \sigma }
\!\!=\!\!
\sqrt{ |\tilde \Delta_{\bf k \sigma}|^2 \!+\! (t_0 \!+ \!t \zeta_{\bf k})^2 }.
\end{eqnarray}

The grand potential of the system is equal to
\begin{eqnarray}
\Omega= \sum_{v = 1}^{4}\sum_{ {\bf k} \sigma}
	(E^{(v)}_{{\bf k} \sigma} - \mu)\,
	\Theta( \mu - E^{(v)}_{{\bf k} \sigma})+B\,,
\end{eqnarray}
where
$\Theta(E)$
is the step-function. Minimization of $\Omega$ with respect to $\langle\gamma^\dag_{\mathbf{p}3\bar{\sigma}}\gamma_{\mathbf{p}2\sigma}^{\phantom{\dag}}\rangle$ and $\langle\gamma^\dag_{\mathbf{p}4\bar{\sigma}}\gamma_{\mathbf{p}1\sigma}^{\phantom{\dag}}\rangle$ gives us the system of equations for $\tilde\Delta_{\bf k \sigma}$ and $\Delta_{\bf k \sigma}$:
\begin{eqnarray}
\label{d1}
\Delta_{\mathbf{k}\sigma}
=
\sum_{\mathbf{p}}
	\left\{
 		\frac{ V^{(1)*}_{\bf pk} \Delta_{{\bf p} \sigma}}
			{2 N_c E^{\rm m}_{{\bf p} \sigma}}
		\left[
			\Theta ( \mu + E^{\rm m}_{{\bf p} \sigma})
			-
			\Theta ( \mu - E^{\rm m}_{{\bf p} \sigma})
		\right]
\right.
\nonumber\\
\left.
		+
 		\frac{ V^{(3)}_{\bf pk} \tilde \Delta_{{\bf p} \sigma}}
			{2 N_c E^{\rm h}_{{\bf p} \sigma}}
		\left[
			\Theta ( \mu + E^{\rm h}_{{\bf p} \sigma})
			-
			\Theta ( \mu - E^{\rm h}_{{\bf p} \sigma})
		\right]
	\right\}\!,\quad
\\
\label{d2}
\tilde \Delta_{\mathbf{k} \sigma}
=
\sum_{\mathbf{p}}
	\left\{
 		\frac{V^{(1)}_{\bf pk} \tilde \Delta_{{\bf p} \sigma}}
			{2 N_c E^{\rm h}_{{\bf p} \sigma}}
		\left[
			\Theta ( \mu + E^{\rm h}_{{\bf p} \sigma})
			-
			\Theta ( \mu - E^{\rm h}_{{\bf p} \sigma})
		\right]
\right.
\nonumber
\\
\left.
		+
 		\frac{V^{(3)}_{\bf pk} \Delta_{{\bf p} \sigma}}
			{2 N_c E^{\rm m}_{{\bf p} \sigma}}
		\left[
			\Theta ( \mu + E^{\rm m}_{{\bf p} \sigma})
			-
			\Theta ( \mu - E^{\rm m}_{{\bf p} \sigma})
		\right]
	\right\}\!.\quad
\end{eqnarray}
The summation in Eqs.~(\ref{d1},\ref{d2}) covers the whole Brillouin zone. However, the interaction $V^{(1,3)}_{\mathbf{pk}}$ is strongest when ${\bf p} \approx {\bf k}$, and decays for larger $|{\bf p} - {\bf k}|$. In the limit of vanishing backscattering
\begin{eqnarray}
\label{bs_def}
V_{\rm bs}^{(1,3)} \equiv V^{(1,3)}_{{\bf K}_1, {\bf K}_2} \approx 0,
\end{eqnarray}
it is possible to define order parameters localized near the specific Dirac point ${\bf K}_\xi$:
$\Delta_{{\bf k} \sigma}=\Delta_{{\bf k}\xi\sigma}$, when ${\bf k} \approx {\bf K}_\xi$.
We see that, within our approximations, the electronic states and the order parameters can be split into four independent sectors, labeled by the multi-index $s=(\sigma,\,\xi)$. A sector with label
$s=(\sigma,\,\xi)$ contains electron states with spin $\sigma$ from valley $\xi$, and hole
states with spin $-\sigma$ from the same valley. This definition implies that all states within a
sector have the same value of the product
$\sigma\xi$.
The sectors are weakly coupled by neglected contributions proportional to
$V_{\rm bs}$
and
$V^{(2,4)}$.
These corrections will be studied perturbatively.

We add and subtract
Eqs.~\eqref{d1}
and~\eqref{d2},
use
Eqs.~\eqref{matrix_V13},
and change the summation by integration over the momentum near the Dirac point $\mathbf{K}_{\xi}$. We also assume that both
$\Delta$ and
$\tilde \Delta$
depend on
$|{\bf k}|$
only. Finally, using the symmetry of our theory with respect to the sign
of $\mu$, we derive for
$0< \mu < t_0$
\begin{eqnarray}
\label{eq::sc_angle_a}
\nonumber
\Delta_{k s}
+
\tilde \Delta_{k s}
=\!
\int_p
	\overline{V}\!(k,p)\!
	\left[
 		\frac{  \Delta_{p s}}
			{2 E^{\rm m}_{p s}}
			\Theta ( E^{\rm m}_{p s} - \mu )
		+
 		\frac{ \tilde \Delta_{p s}}
			{2 E^{\rm h}_{p s}}
	\right]\!,
\\
\Delta_{k s}
-
\tilde \Delta_{{k} s}
=\!
\int_p
	\overline{U}\!(k,p)\!
	\left[
 		\frac{  \Delta_{p s}}
			{2 E^{\rm m}_{p s}}
			\Theta ( E^{\rm m}_{p s} - \mu )
		-
 		\frac{ \tilde \Delta_{p s}}
			{2 E^{\rm h}_{p s}}
	\right]\!,
\end{eqnarray}
where $\int_p \ldots = (2 \pi  /v_{\rm BZ})\int  pdp \ldots$, and the volume (area) of the Brillouin zone is $v_{\rm BZ} = 8\pi^2/(3\sqrt{3}a_0^2)$. In Eqs.~(\ref{eq::sc_angle_a}), the averaged coupling constants are
\begin{eqnarray}
\label{eq::V01_def}
\overline{V} (k,p)
&=&
\int \frac{d \phi}{2\pi}
V_C (\sqrt{k^2 + p^2 - 2 k p \cos \phi}),
\\
\nonumber
\overline{U} (k,p)
&=&
\int \frac{d \phi}{2\pi}
V_C (\sqrt{k^2 + p^2 - 2 k p \cos \phi}) \cos \phi,
\end{eqnarray}
and the
spectrum~\eqref{spectrum}
in sector
$s=(\sigma, \xi)$
can be approximated as
\begin{eqnarray}
\label{spectrum_Dec}
E^{\rm m}_{p s}
&\cong&
\sqrt{|\Delta_{s}|^2 + t_0^2(1 - p/k_{F0})^2},
\\
\nonumber
E^{\rm h}_{p s}
&\cong&
\sqrt{|\tilde{\Delta}_{s}|^2 + t_0^2(1 + p/k_{F0})^2}
\cong
t_0(1 + p/k_{F0}),
\quad
\quad
\end{eqnarray}
where
$p=|\mathbf{p}-\mathbf{K}_\xi|$.

{\it BCS-like approximation.}--- In general, we can choose some model for
$V_C(q)$ and solve Eqs.~\eqref{eq::sc_angle_a}
numerically. However, modeling the effective Coulomb interaction in graphene
bilayers is notoriously difficult, and no universal and compact answer is
known~\cite{note_coulomb}. In this situation, finding an accurate numerical solution to the integral equations~\eqref{eq::sc_angle_a} is impractical. Instead, we use the simple BCS-like
ansatz
$\Delta_s(q)=\Delta_s \Theta(\Lambda - |q - k_{F0}|)$
and
$\tilde{\Delta}_s(q)
=
\tilde{\Delta}_s \Theta(\Lambda - |q - k_{F0}|)$
for the order parameters (the cutoff momentum $\Lambda$ satisfies
$\Lambda<k_{F0}$),
and assume that
$\overline{V}$
and $\overline{U}$ are constants independent of $k$ and $p$. We believe that this ansatz, despite its simplicity, captures all the necessary physics. Now the integral equations become non-linear algebraic equations
\begin{eqnarray}
\label{eq::MF_eqs}
\Delta_{s} + \tilde \Delta_{ s}
=
g \Delta_s \ln \left(\frac{E^*}{\mu+\sqrt{\mu^2-\Delta_s^2}} \right)
+
\tilde g \tilde \Delta_s,\nonumber\\
\Delta_{s} - \tilde \Delta_{ s}
=
\frac{g}{\alpha} \Delta_s \ln \left(
	\frac{E^*}{\mu+\sqrt{\mu^2-\Delta_s^2}}
\right)
-
\frac{\tilde g}{\alpha} \tilde \Delta_s,
\end{eqnarray}
where the energy scale is
$E^* = 2 t_0 \Lambda / k_{F0}$ and
the coupling constants are
\begin{eqnarray}
\label{gg}
g = \frac{t_0}{\sqrt{3}\pi t^2} \overline{V},
\quad
\tilde g = \frac{\Lambda}{2 k_{F0}} g,
\quad
\alpha = {\overline{V}}/{\overline{U}}>1.
\end{eqnarray}
It trivially follows from Eqs.~(\ref{eq::MF_eqs}) that $\tilde \Delta_s = C \Delta_s$,
where
$C = (\alpha - 1)/(\alpha + 1 - 2 \tilde g) $.
At zero doping, which corresponds to the case
$\mu=\Delta_s$,
one finds
\begin{equation}
\label{eq::Delta0}
\Delta_s = \Delta_0 = E^* \exp{\left[-\frac{1}{g}\frac{2\alpha-\tilde{g}(1+\alpha)}{1+\alpha-2\tilde{g}}\right]}.
\end{equation}
This compact mean field solution is valid in the small coupling limit; that is,
when $g$ (and $\tilde{g}$) is small, and, consequently,
$\Delta_0$ and $\tilde \Delta_0$
are much less than
$t_0$. The doped state is characterized by
$\mu > \Delta_s$. To describe the solution of
Eq.~(\ref{eq::MF_eqs})
in such a regime, let us define the partial doping
$x_s$:
the concentration of electrons residing in sector $s$, per single carbon
atom. It is
known~\cite{Rice,our_chrom2013,Sboychakov_PRB2013_PS_pnict,prb_sl2017}
that a finite $x_s$ acts to decrease the order parameter
$\Delta_s$:
\begin{equation}
\label{DeltaDop}
\Delta_s (x_s) =\Delta_0\sqrt{1-\frac{4x_s}{x_0}},
\quad
\mu=\Delta_0\left(1-\frac{2x_s}{x_0}\right).
\end{equation}
where
$x_0=\Delta_0t_0/(\pi\sqrt{3}t^2)$.
It is easy to check that
Eqs.~(\ref{DeltaDop})
indeed guarantee that $\mu$ exceeds
$\Delta_s$,
making the doping of sector $s$ possible.
At $T=0$ the
partial free energy (per unit cell) associated with doping is
\begin{eqnarray}
\label{Free_en}
\Delta F_{s}(x_{s}) = 4\int_0^{x_{s}}\!\!\!\!\!\mu(x)dx
= 4 \Delta_0 \left( x_s -  \frac{x_s^2}{x_0}\right).
\end{eqnarray}
Since a unit cell contains four carbon atoms, the factor 4 is required in
this formula.

{\it Fractional metal state.}--- The relations~(\ref{DeltaDop},\ref{Free_en}) describe a single sector. To determine the state of the whole system, we
must understand how the total doping $x$ is distributed between the
sectors. One might expect that $x$ is spread evenly:
$x_s = x/4$.
Yet such an assumption might not be most advantageous thermodynamically: we
demonstrated~\cite{our_hmet_prl2017,our_hmet_prb2018}, for a two-sector system, that placing all the extra charge $x$ into a single sector optimizes the system free energy relative to the state with an even distribution of $x$. To settle this issue for our four-sector model, we must
minimize the doping-related part of the free energy for the whole system
\begin{equation}
\label{FreeEnergy_tot}
\Delta F= \sum_s \Delta F_s = 4 \Delta_0 x-
	\frac{4 \Delta_0}{x_0}\sum_{\xi\sigma}{x_{\xi\sigma}^2}
\end{equation}
at fixed doping $x = \sum_s x_s$. Simple calculations demonstrate that, for
$x<x_0$,
the term
$\Delta F$
reaches its smallest value,
$\Delta F_{\rm qm} = 4 \Delta_0 (x - x^2/x_0)$,
when all extra electrons are placed into a specific sector $s$, while all
other sectors are kept doping-free
\begin{eqnarray}
\label{eq::quarter_met}
x_s = x,
\quad
x_{s'} = 0
\quad
\text{for}
\quad
s' \ne s.
\end{eqnarray}
For example, $\Delta F_{\rm qm}$
is smaller than
$\Delta F_{\rm e} = 4 \Delta_0 x -\Delta_0x^2/x_0$,
which is the free energy of the state with
$x_s = x/4$ for all four $s$. For the distribution~(\ref{eq::quarter_met})
the Fermi surface lies entirely in sector
$s=(\sigma, \xi)$.
Therefore, only states with spin $\sigma$ near the Dirac point
${\bf K}_\xi$
reach the Fermi level. In other words, the Fermi surface is perfectly
polarized in terms of both $\sigma$ and $\xi$ indices. Since the insulating
gap persists in three other sectors, the state described by
Eq.~(\ref{eq::quarter_met})
may be called `a quarter-metal', a first example of a series of `fractional
metals'.

As in the case of the half-metal in the system with nesting~\cite{our_hmet_prl2017,our_hmet_prb2018}, the gap in the first sector closes when increasing doping. The doped electrons begin to enter the second sector, then to the third and fourth sectors. As a result, the system passes respectively through the states of a half-metal, 3/4-metal, and finally the gaps in all sectors close and the system occurs in the usual metallic phase. We can show that each transformation is a first-order phase transition. The analysis of the electronic states evolution with doping is quite similar to the half-metal case~\cite{our_hmet_prl2017,our_hmet_prb2018}.

{\it Stability of Fractional metal.}--- Above we neglected interactions between electrons in different sectors. Then, treating individual sectors independently, we derived Eqs.~(\ref{DeltaDop},\ref{Free_en}). Now we want to assess the effects of the neglected terms. There are two types of interaction terms: (i)~umklapp interaction $\hat{H}^{(2,4)}$, Eqs.~(\ref{H2},\ref{H4}), which couples sectors with the same $\xi$ but different spins, and (ii)~the backscattering amplitude $V_{\rm bs}^{(1,3)}$, which describes interactions between sectors with the same $\sigma$ but different valley $\xi$, Eq.~\eqref{bs_def}. In principle, $\hat{H}^{(2,4)}$ also contain the backscattering $V_{\rm bs}^{(2)}$, which is even weaker, and will be neglected. If the associated coupling constants are small, we can use perturbation theory. The lowest-order perturbative correction
$F_{\rm um}$ to the free energy due to the umklapp term
$\hat{H}^{(2)}$ equals $\langle \hat{H}^{(2)} \rangle$.
Thus, neglecting small contributions due to
$\tilde \Delta_s$,
we determine the umklapp correction to the free energy (per unit cell)
\begin{eqnarray}
F_{\rm um}
= - \frac{\cal F}{2}
\sum_\xi
	\sqrt{\left(1-\frac{4 x_{\uparrow \xi}}{x_0} \right)
	\left(1-\frac{4 x_{\downarrow \xi}}{x_0}\right)},
\end{eqnarray}
where ${\cal F} = 8\alpha^2 g_{\rm um}\Delta_0 x_0/{(1+\alpha)^2 g^2}$, and the dimensionless Fermi-surface-averaged umklapp coupling constant is $g_{\rm um}=t_0\overline{V}_{\rm um}/\sqrt{3}\pi t^2$. We also used the fact that $V^{(1)}$, upon averaging over the Fermi surface, becomes equal to $g (1 + \alpha)/2 \alpha$.
When $x$ is low, one has
$F_{\rm um} /{\cal F}
\approx
	- 1 + x/x_0
	+
	\sum_{\xi} {(x_{\uparrow \xi} - x_{\downarrow \xi})^2}/{x_0^2}$,
which is smallest at
$x_s = x/4$.
A similar result can be derived for the backscattering interaction. Thus,
both the umklapp and the backscattering favor an even distribution of doping
over the sectors. However, in the limit
$g_{\rm um} \ll g^2$,
$g_{\rm bs} \ll g^2$,
their contributions are small, and cannot destroy the fractional metal
phase. The perturbative derivation of the stability criterion is intuitively clear
and transparent. Its primary purpose is to demonstrate that the fractional
metal phase can survive weak deviations from the highly idealized
model neglecting any couplings between the sectors. On the other hand, this
criterion is very stringent, and one may wonder if it can be satisfied in a
real material. Fortunately, a more complex non-perturbative approach, which
accounts for the inter-sector couplings at the mean field level, allows to
relax it: we demonstrated~\cite{supplemental}
that it is sufficient to have
\begin{equation}
\label{stab_condition}
g_{\rm bs} < g,\quad g_{\rm um} < g
\end{equation}
to maintain the stability of the FraM. More detailed stability analysis
will be presented in future studies.

\textit{Discussion.}--- Using AA~bilayer graphene as a test example, we
argue that in a system with a nested multi-sheet Fermi surface, a peculiar
state (which we call fractional metal, or FraM) can be stabilized.
In the FraM phase, part of the Fermi surface is gapped and charge carriers
on the remaining gapless part of the Fermi surface belong to a specific sector
of the low-energy electronic states. Similar to a half-metal, the states at
the Fermi energy can be characterized in terms of polarization; but, unlike
the usual half-metals, this is not spin polarization. Let us introduce the spin-flavor~\cite{note_on_spin_flavor} operator $\hat{S}_{\rm f}=\sum_{\sigma\xi\nu}\sigma \nu \hat{N}_{\sigma\xi\nu}$, where $\hat{N}_{\sigma\xi \nu}$
is the number operator for fermions with spin $\sigma$, charge $\nu$, in
valley $\xi$. Since doping enters only in one sector, all states at the
Fermi surface have the same value of $\sigma \nu$. Therefore, these states are eigenstates of
$\hat{S}_{\rm f}$ with the same eigenvalue $\sigma \nu$. The same is true for the valley operator
$\hat{S}_{\rm v}=\sum_{\sigma\xi\nu} \xi \hat{N}_{\sigma\xi\nu}$,
since a given sector is localized entirely in one valley.

Thus, \textit{the Fermi surface of the FraM is polarized in terms of two spin-like operators $S_{\rm f,v}$.} This implies that \textit{the electric current though the FraM carries, in addition to the electric charge, spin-flavor and valley quanta.}  Finally, note that, if superconductivity arises in a FraM phase, it should obey rather peculiar properties. The superconducting order parameter might have a very unusual symmetry, classified according to a non-trivial spin and valley structure, and superconducting currents would be spin-flavor and valley polarized. However, the detailed analysis of this superconductivity requires the specification of the symmetric properties of the electron-phonon coupling.

\section*{Acknowledgment}

This work is partially supported by the Russian Foundation for Basic Research (RFBR) under grant no. 19-02-00421 and JSPS-RFBR program under grant no. 19-52-50015. F.N. is supported in part by: NTT Research, Army Research Office (ARO) (Grant No. W911NF-18-1-0358), Japan Science and Technology Agency (JST) (via the CREST Grant No. JPMJCR1676), Japan Society for the Promotion of Science (JSPS) (via the KAKENHI Grant No. JP20H00134 and the JSPS-RFBR Grant No. JPJSBP120194828), the Asian Office of Aerospace Research and Development (AOARD), and the Foundational Questions Institute Fund (FQXi) via Grant No. FQXi-IAF19-06.

\begin{widetext}

\

\

\section{Appendix: Can bilayer graphene become a fractional metal?}

Below we show the study of the stability of the quarter-metal against the
umklapp interaction term.

\section{Basic equations}

For reader's convenience, let us recall several basic equations and facts
from the main text. 

\subsection{Definitions}

The interaction Hamiltonian is
\begin{eqnarray}
\label{suppl::eq::int_gamma_subst}
\hat{H}_{\textrm{int}}
&=& \hat{H}^{(1)} + \hat{H}^{(2)} + \hat{H}^{(3)} + \hat{H}^{(4)},
\quad
\text{where}
\quad
\\
\hat{H}^{(1)}
&=& 
	-\frac{1}{N_c}
	\sum_{\mathbf{kp}  \sigma }
		V^{(1)}_{ {\bf k} , {\bf p} } \left[
			(\gamma_{\mathbf{k}1\sigma}^\dag
			\gamma_{\mathbf{k}4\bar{\sigma}})
			(\gamma_{\mathbf{p}4\bar{\sigma}}^\dag
			\gamma_{\mathbf{p}1\sigma})
			+
			(\gamma_{\mathbf{k}3\bar{\sigma}}^\dag
			\gamma_{\mathbf{k}2\sigma})
			(\gamma_{\mathbf{p}2\sigma}^\dag
			\gamma_{\mathbf{p}3\bar{\sigma}})
	\right],
\\
\hat{H}^{(2)}
&=&
	-\frac{1}{2N_c}
	\sum_{\mathbf{kp}  \sigma }
		V^{(2)}_{ {\bf k} , {\bf p} } \left[
			(\gamma_{\mathbf{k}1\sigma}^\dag
			\gamma_{\mathbf{k}4\bar{\sigma}})
			(\gamma_{\mathbf{p}1\bar{\sigma}}^\dag
			\gamma_{\mathbf{p}4\sigma})
			+
			(\gamma_{\mathbf{k}2\sigma}^\dag
			\gamma_{\mathbf{k}3\bar{\sigma}})
			(\gamma_{\mathbf{p}2\bar{\sigma}}^\dag
			\gamma_{\mathbf{p}3\sigma})
			+
			{\rm H.c.}
	\right],
\\
\hat{H}^{(3)}
&=&
	-\frac{1}{N_c}
	\sum_{\mathbf{kp}  \sigma }
		V^{(3)}_{ {\bf k} , {\bf p} } \left[
			(\gamma_{\mathbf{k}1\sigma}^\dag
			\gamma_{\mathbf{k}4\bar{\sigma}})
			(\gamma_{\mathbf{p}3\bar{\sigma}}^\dag
			\gamma_{\mathbf{p}2\sigma})
			+
			(\gamma_{\mathbf{k}2\sigma}^\dag
			\gamma_{\mathbf{k}3\bar{\sigma}})
			(\gamma_{\mathbf{p}4\bar{\sigma}}^\dag
			\gamma_{\mathbf{p}1\sigma})
	\right],
\\
\hat{H}^{(4)}
&=&
	-\frac{1}{2N_c}
	\sum_{\mathbf{kp}  \sigma }
		V^{(4)}_{ {\bf k} , {\bf p} } \left[
			(\gamma_{\mathbf{k}1\sigma}^\dag
			\gamma_{\mathbf{k}4\bar{\sigma}})
			(\gamma_{\mathbf{p}2\bar{\sigma}}^\dag
			\gamma_{\mathbf{p}3\sigma})
			+
			(\gamma_{\mathbf{k}2\sigma}^\dag
			\gamma_{\mathbf{k}3\bar{\sigma}})
			(\gamma_{\mathbf{p}1\bar{\sigma}}^\dag
			\gamma_{\mathbf{p}4\sigma})
			+
			{\rm H.c.}
		\right],
\end{eqnarray}
with the coupling constants
$V^{(1,2,3,4)}_{\bf k,p}$
defined as
\begin{eqnarray}
\label{suppl::V13}
V^{(1,3)}_{\bf k,p} = \frac{1}{8}
\left[
	V^{00}_{AA}\!+\!V^{10}_{AA}
	\!\pm\!
	\left(V^{00}_{AB}\!+\!V^{10}_{AB}\right)e^{-i\Delta\varphi}
	\!+\!
	{\rm C.c.}
\right]\!,
\quad
\\
\label{suppl::V24}
V^{(2,4)}_{\bf k,p} = \frac{1}{8}
\left[
	V^{00}_{AA}\!-\!V^{10}_{AA}\!
	\mp\!
	\left(V^{00}_{AB}\!-\!V^{10}_{AB}\right)e^{-i\Delta\varphi}
	\!+\!
	{\rm C.c.}
\right]\!.
\quad
\end{eqnarray}
Our first-step approximation is
\begin{eqnarray}
\label{suppl::matrix_V13}
V^{(1,3)}_{\bf k,p}
\approx
\frac{1}{2} V_C(\mathbf{k-p})
	\left[1\pm \cos (\Delta\varphi_\mathbf{k,p})\right],
\quad
V^{(2,4)}_{\bf k,p} \approx 0.
\quad
\end{eqnarray}
The interaction can be approximated as
$\hat{H}_{\textrm{int}} \approx \hat{H}^{(1)} + \hat{H}^{(3)}$.

\subsection{Mean field approximation}

The mean field version of 
$\hat{H}_{\textrm{int}}$
is
\begin{eqnarray}
\label{suppl::eq::int_MF_fin}
\hat{H}_{\textrm{int}}^{\rm MF}
=
\frac{1}{N_c}
\left(
	B_\uparrow\! +\!\! B_\downarrow
\right)
-\sum_{\mathbf{p}  \sigma }
	\left(
		\tilde \Delta_{{\bf p} \sigma}
		\gamma_{\mathbf{p}4\bar{\sigma}}^\dag
		\gamma_{\mathbf{p}1\sigma}
		\!\!+\!\!
		\Delta_{{\bf p} \sigma}
		\gamma_{\mathbf{p}3\bar{\sigma}}^\dag
		\gamma_{\mathbf{p}2\sigma}
		\!+\!
		{\rm H.c.}
	\!\right)\!\!,\,
\end{eqnarray}
where
\begin{eqnarray}
\label{suppl::Delta}
\Delta_{\mathbf{k} \sigma}
&=& 
\frac{1}{N_c}
\sum_{\mathbf{p}}
	\!\left[
		V^{(1)}_\mathbf{p,k}
		\langle
			\gamma^\dag_{\mathbf{p}2\sigma}
			\gamma_{\mathbf{p}3\bar{\sigma}}
		\rangle
		\!+\!
  		V^{(3)}_{\mathbf{p,k}}
		\langle
			\gamma^\dag_{\mathbf{p}1\sigma}
			\gamma_{\mathbf{p}4\bar{\sigma}}
		\rangle
	\right],
\\
\tilde{\Delta}_{\mathbf{k} \sigma}
&=& 
\frac{1}{N_c}\!
\sum_{\mathbf{p}}
  	\left[
		V^{(1)}_{\mathbf{p,k}}
		\langle
			\gamma^\dag_{\mathbf{p}1\sigma}
			\gamma_{\mathbf{p}4\bar{\sigma}}
		\rangle
		\!+\!
  		V^{(3)}_{\mathbf{p,k}}
		\langle
			\gamma^\dag_{\mathbf{p}2\sigma}
			\gamma_{\mathbf{p}3\bar{\sigma}}
		\rangle
	\right],
\\
B_\sigma
&=& 
\frac{1}{N_c}
\sum_{\mathbf{k}}
	\left[
		\Delta_{\mathbf{k} \sigma}
		\langle
			\gamma^\dag_{\mathbf{k}3\bar{\sigma}}
			\gamma_{\mathbf{k}2\sigma}
		\rangle
  		\!+\!
		\tilde{\Delta}_{\mathbf{k} \sigma}
		\langle
			\gamma^\dag_{\mathbf{k}4\bar{\sigma}}
			\gamma_{\mathbf{k}1\sigma}
		\rangle
	\right].
\end{eqnarray}
The spectrum of the mean-field Hamiltonian can be easily derived
\begin{eqnarray}
\label{suppl::spectrum}
\!\!\!\!&&E^{(2,3)}_{{\bf k} \sigma} = \mp E^{\rm m}_{\bf k \sigma },
\qquad
E^{(1,4)}_{{\bf k} \sigma} = \mp E^{\rm h}_{\bf k \sigma },
\end{eqnarray} 
where
\begin{eqnarray} 
\nonumber
\!\!\!\!&&E^{\rm m}_{\bf k \sigma }\!\!=\!\!\sqrt{ |\Delta_{\bf k \sigma}|^2\! +\! (t_0\! -\! t \zeta_{\bf k})^2 },\,\,
\quad
E^{\rm h}_{\bf k \sigma }
\!\!=\!\!
\sqrt{ |\tilde \Delta_{\bf k \sigma}|^2 \!+\! (t_0 \!+ \!t \zeta_{\bf k})^2 }.
\end{eqnarray}
The total energy of the system is 
\begin{eqnarray}
E = \sum_{\nu = 1}^{4}\sum_{ {\bf k} \sigma}
	(E^{(\nu)}_{{\bf k} \sigma} - \mu)\,
	\Theta( \mu - E^{(\nu)}_{{\bf k} \sigma}),
\end{eqnarray}
where
$\Theta(E)$
is the step-function. Using the Hellmann-Feynman theorem, we obtain
\begin{eqnarray}
\label{suppl::averages}
\nonumber
\langle
	\gamma^\dag_{\mathbf{k}3\bar{\sigma}}
	\gamma_{\mathbf{k}2\sigma}
\rangle
=
 \frac{ \Delta_{{\bf k} \sigma}}{2 E^{\rm m}_{{\bf k} \sigma}}
\left[ \Theta ( \mu + E^{\rm m}_{{\bf k} \sigma})
	- \Theta ( \mu - E^{\rm m}_{{\bf k} \sigma})
\right],\\
\langle
	\gamma^\dag_{\mathbf{k}4\bar{\sigma}}
	\gamma_{\mathbf{k}1\sigma}
\rangle
=
 \frac{ \tilde \Delta_{{\bf k} \sigma}}{2 E^{\rm h}_{{\bf k} \sigma}}
\left[ \Theta ( \mu + E^{\rm h}_{{\bf k} \sigma})
	- \Theta ( \mu - E^{\rm h}_{{\bf k} \sigma})
\right].
\end{eqnarray}
Formally, the summation in
Eq.~\eqref{suppl::Delta}
covers the whole Brillouin zone. However, the interaction 
$V^{(1,3)}_{\mathbf{p,k}}$
is the strongest when
${\bf p} \approx {\bf k}$,
and decays for larger
$|{\bf p} - {\bf k}|$.
In the limit of vanishing backscattering
\begin{eqnarray}
\label{suppl::bs_def}
V_{\rm bs}^{(1,3)} \equiv V^{(1,3)}_{{\bf K}_1, {\bf K}_2} \approx 0,
\end{eqnarray} 
it is possible to define order parameters localized near a specific Dirac
point
${\bf K}_\xi$:
$\Delta_{{\bf k} \sigma \xi} = \Delta_{{\bf k} \sigma}$,
when 
${\bf k} \approx {\bf K}_\xi$.
Combining
Eqs.~\eqref{suppl::Delta}
and~\eqref{suppl::averages},
we obtain the self-consistent equations in the form
\begin{eqnarray}
\Delta_{\mathbf{k} \sigma \xi}
=
\frac{1}{N_c}
\sum_{\mathbf{p} \in {\bf K}_\xi}
	\left\{
 		\frac{ V^{(1)}_{\bf p,k} \Delta_{{\bf p} \sigma \xi}}
			{2 E^{\rm m}_{{\bf p} \sigma}}
		\left[
			\Theta ( \mu + E^{\rm m}_{{\bf p} \sigma})
			-
			\Theta ( \mu - E^{\rm m}_{{\bf p} \sigma})
		\right]
		+
 		\frac{V^{(3)}_{\bf p,k} \tilde \Delta_{{\bf p}\sigma \xi}}
			{2 E^{\rm h}_{{\bf p} \sigma}}
		\left[
			\Theta ( \mu + E^{\rm h}_{{\bf p} \sigma})
			-
			\Theta ( \mu - E^{\rm h}_{{\bf p} \sigma})
		\right]
	\right\},
\\
\tilde \Delta_{\mathbf{k} \sigma \xi}
=
\frac{1}{N_c}
\sum_{\mathbf{p} \in {\bf K}_\xi}
	\left\{
 		\frac{ V^{(1)} \tilde \Delta_{{\bf p} \sigma \xi}}
			{2 E^{\rm h}_{{\bf p} \sigma}}
		\left[
			\Theta ( \mu + E^{\rm h}_{{\bf p} \sigma})
			-
			\Theta ( \mu - E^{\rm h}_{{\bf p} \sigma})
		\right]
		+
 		\frac{ V^{(3)} \Delta_{{\bf p} \sigma \xi}}
			{2 E^{\rm m}_{{\bf p} \sigma}}
		\left[
			\Theta ( \mu + E^{\rm m}_{{\bf p} \sigma})
			-
			\Theta ( \mu - E^{\rm m}_{{\bf p} \sigma})
		\right]
	\right\}.
\end{eqnarray}
Simplifying the latter equations in the regime
$\mu > 0$,
we derive
\begin{eqnarray}
\label{suppl::d1}
\Delta_{\mathbf{k} \sigma \xi}
=
\frac{1}{N_c}
\sum_{\mathbf{p} \in {\bf K}_\xi}
	\left\{
 		\frac{ V^{(1)}_{\bf p,k} \Delta_{{\bf p} \sigma \xi}}
			{2 E^{\rm m}_{{\bf p} \sigma}}
			\Theta ( E^{\rm m}_{{\bf p} \sigma} - \mu )
		+
 		\frac{V^{(3)}_{\bf p,k} \tilde \Delta_{{\bf p}\sigma \xi}}
			{2 E^{\rm h}_{{\bf p} \sigma}}
			\Theta ( E^{\rm h}_{{\bf p} \sigma} - \mu )
	\right\},
\\
\label{suppl::d2}
\tilde \Delta_{\mathbf{k} \sigma \xi}
=
\frac{1}{N_c}
\sum_{\mathbf{p} \in {\bf K}_\xi}
	\left\{
 		\frac{ V^{(1)} \tilde \Delta_{{\bf p} \sigma \xi}}
			{2 E^{\rm h}_{{\bf p} \sigma}}
			\Theta ( E^{\rm h}_{{\bf p} \sigma} - \mu )
		+
 		\frac{ V^{(3)} \Delta_{{\bf p} \sigma \xi}}
			{2 E^{\rm m}_{{\bf p} \sigma}}
			\Theta ( E^{\rm m}_{{\bf p} \sigma} - \mu )
	\right\}.
\end{eqnarray}
We see that, within our approximations, the electronic states and the order
parameters can be split into four independent sectors, which can be labeled
by the multi-index
$s=(\sigma,\xi)$.
Our derivation implies that the sectors are not entirely independent:
neglected contributions proportional to
$V_{\rm bs}$
and
$V^{(2,4)}$
couple them. Due to the smallness of these couplings, they can be treated
perturbatively.

We add and subtract
Eqs.~\eqref{suppl::d1}
and~\eqref{suppl::d2},
use
Eqs.~\eqref{suppl::matrix_V13},
and change the summation over momentum by an integration. We also assume
that both $\Delta$ and
$\tilde \Delta$
only depend on
$|{\bf k}|$.
Finally, using the symmetry of our theory with respect to the sign
of $\mu$, we derive for 
$0< \mu < t_0$
\begin{eqnarray}
\label{suppl::eq::sc_angle_a}
\nonumber 
\Delta_{k s}
+
\tilde \Delta_{k s}
=\!
\int_p
	\bar{V}^{(0)}_Q \!
	\left[
 		\frac{  \Delta_{p s}}
			{2 E^{\rm m}_{p s}}
			\Theta ( E^{\rm m}_{p s} - \mu )
		+
 		\frac{ \tilde \Delta_{p s}}
			{2 E^{\rm h}_{p s}}
	\right]\!,
\\
\Delta_{k s}
-
\tilde \Delta_{{k} s}
=\!
\int_p
	\bar{V}^{(1)}_Q \!
	\left[
 		\frac{  \Delta_{p s}}
			{2 E^{\rm m}_{p s}}
			\Theta ( E^{\rm m}_{p s} - \mu )
		-
 		\frac{ \tilde \Delta_{p s}}
			{2 E^{\rm h}_{p s}}
	\right]\!,
\end{eqnarray}
where the integration symbol stands for
$\int_p \ldots = (2 \pi p /v_{\rm BZ})\int  dp \ldots$,
and the volume (area) of the Brillouin zone is
$v_{\rm BZ} = 8\pi^2/(3\sqrt{3}a^2)$.
In 
Eqs.~(\ref{suppl::eq::sc_angle_a}),
the averaged coupling constants are
\begin{eqnarray}
\label{suppl::eq::V01_def}
\bar{V}^{(0)}_Q (k,p)
&=&
\int \frac{d \phi}{2\pi}
V_C (\sqrt{k^2 + p^2 + 2 k p \cos \phi}),
\\
\nonumber
\bar{V}^{(1)}_Q (k,p)
&=&
\int \frac{d \phi}{2\pi}
V_C (\sqrt{k^2 + p^2 + 2 k p \cos \phi}) \cos \phi,
\end{eqnarray}
and the 
spectrum~\eqref{suppl::spectrum}
in sector 
$s=(\sigma, \xi)$
can be approximated as
\begin{eqnarray}
\label{suppl::spectrum_Dec}
E^{\rm m}_{p s}
&\approx&
\sqrt{|\Delta_{s}|^2 + t_0^2(1 - p/k_{F0})^2},
\\
\nonumber
E^{\rm h}_{p s}
&\approx&
\sqrt{|\tilde{\Delta}_{s}|^2 + t_0^2(1 + p/k_{F0})^2}
\approx
t_0(1 + p/k_{F0}),
\quad
\quad
\end{eqnarray}
where
$p=|\mathbf{p}-\mathbf{K}_\xi|$.

To solve the integral
equations~(\ref{suppl::eq::sc_angle_a})
we use the simple BCS-like
ansatz
\begin{eqnarray} 
\Delta_a (q)=\Delta_s \Theta(\Lambda - |q - k_{F0}|)
\quad
\text{and}
\quad
\tilde{\Delta}_{\xi\sigma}(q)
=
\tilde{\Delta}_s \Theta(\Lambda - |q - k_{F0}|)
\end{eqnarray} 
for the order parameters (the cutoff momentum $\Lambda$ satisfies
$\Lambda \ll Q_0$),
and assume that
$\bar{V}^{(0,1)}_Q$
are constants independent of $k$ and $p$. This allows us to convert the
integral equations into non-linear algebraic equations
\begin{eqnarray} 
\label{suppl::eq::MF_eqs}
\Delta_{s} + \tilde \Delta_{ s}
=
g \Delta_s \ln \left(\frac{E^*}{\mu+\sqrt{\mu^2-\Delta_s^2}} \right) 
+
\tilde g \tilde \Delta_s,
\\
\nonumber 
\Delta_{s} - \tilde \Delta_{ s}
=
\frac{g}{\alpha} \Delta_s \ln \left(
	\frac{E^*}{\mu+\sqrt{\mu^2-\Delta_s^2}} 
\right) 
-
\frac{\tilde g}{\alpha} \tilde \Delta_s,
\end{eqnarray} 
where the energy scale is
$E^* = 2 t_0 \Lambda / k_{F0}$,
and the coupling constants are
\begin{eqnarray}
g = \frac{t_0}{\sqrt{3}\pi t^2} \bar{V}^{(0)},
\quad
\tilde g = \frac{\Lambda}{2 k_{F0}} g,
\quad
\alpha = \frac{\bar{V}^{(0)}}{\bar{V}^{(1)}}>1.
\end{eqnarray} 

\subsection{Solutions of the mean field equations}

At zero doping, which corresponds to the case
$\mu=\Delta_s$,
the order parameters are
\begin{eqnarray}
\label{suppl::eq::Delta0}
\Delta_s = \Delta_0
=
E^* \exp{\left[
	-\frac{1}{g}
	\frac{2\alpha-\tilde{g}(1+\alpha)}{1+\alpha-2\tilde{g}}
\right]},
\qquad
\tilde \Delta_s = \frac{\alpha - 1}{\alpha + 1 - 2 \tilde g}  \Delta_s.
\end{eqnarray} 
This mean-field solution is valid in the weak-coupling limit, that is, when
$g$ is small, and, consequently, 
$\Delta_0$
and 
$\tilde \Delta_0$
are much less than 
$t_0$.
The doped state is characterized by 
$\mu > \Delta_s$.
To describe the solution of
Eq.~(\ref{suppl::eq::MF_eqs})
in such a regime, let us define the partial doping
$x_s$
for the concentration of electrons residing in sector $s$. It is known that
a finite 
$x_s$
decreases the order parameter
$\Delta_s$:
\begin{equation}
\label{suppl::DeltaDop}
\Delta_s (x_s) =\Delta_0\sqrt{1-\frac{4x_s}{x_0}},
\qquad
\mu=\Delta_0\left(1-\frac{2x_s}{x_0}\right),
\end{equation}
where
$x_0=\Delta_0t_0/(\pi\sqrt{3}t^2)$.
It is easy to check that 
Eqs.~(\ref{suppl::DeltaDop}) 
indeed guarantee that $\mu$ exceeds 
$\Delta_s$,
making it possible to dope sector $s$. At zero temperature, the partial
free energy (per unit cell) associated with doping becomes
\begin{eqnarray} 
\label{suppl::Free_en}
\Delta F_{s}(x_{s}) = 4 \int_0^{x_{s}}\!\! \mu(x)\,dx
= 4 \Delta_0 \left( x_s - \frac{x_s^2}{x_0}\right).
\end{eqnarray} 
As in the main text, the factor 4 accounts for the four carbon atoms in a
single unit cell.
 
\section{Stability against the umklapp interaction}

\subsection{Self-consistent equations}

The next step is to add the inter-sector interaction. We will use
$H^{(2)}$
as an example of the inter-sector interaction. The other example is the
backscattering
$V_{\rm bs}^{(1,3)}$.
The term
$H^{(2)}$
is a type of umklapp scattering: such term is non-zero only when the
nesting vector is either zero or half of the elementary reciprocal lattice
vector. If we average
$H^{(2)}$
we obtain
\begin{eqnarray} 
\langle \hat{H}^{(2)} \rangle 
=
	-\frac{1}{2N_c}
	\sum_{\mathbf{kp}  \sigma }
		V^{(2)}_{ {\bf k} , {\bf p} } \left[
			\langle \gamma_{\mathbf{k}1\sigma}^\dag
			\gamma_{\mathbf{k}4\bar{\sigma}} \rangle
			\langle \gamma_{\mathbf{p}1\bar{\sigma}}^\dag
			\gamma_{\mathbf{p}4\sigma} \rangle
			+
			\langle \gamma_{\mathbf{k}2\sigma}^\dag
			\gamma_{\mathbf{k}3\bar{\sigma}} \rangle
			\langle \gamma_{\mathbf{p}2\bar{\sigma}}^\dag
			\gamma_{\mathbf{p}3\sigma} \rangle
			+
			{\rm C.c.}
	\right]
\\
\nonumber
\approx
	-\frac{1}{2N_c}
	\sum_{\mathbf{kp}  \sigma }
		V^{(2)}_{ {\bf k} , {\bf p} } \left[
			\langle \gamma_{\mathbf{k}2\sigma}^\dag
			\gamma_{\mathbf{k}3\bar{\sigma}} \rangle
			\langle \gamma_{\mathbf{p}2\bar{\sigma}}^\dag
			\gamma_{\mathbf{p}3\sigma} \rangle
			+
			{\rm C.c.}
	\right]
\approx
	-\frac{\bar{V}_{\rm um}}{2N_c}
	\sum_{\mathbf{kp}  \sigma } \left[
		\langle \gamma_{\mathbf{k}2\sigma}^\dag
		\gamma_{\mathbf{k}3\bar{\sigma}} \rangle
		\langle \gamma_{\mathbf{p}2\bar{\sigma}}^\dag
		\gamma_{\mathbf{p}3\sigma} \rangle
		+
		{\rm C.c.}
	\right],
\end{eqnarray} 
where 
$\bar{V}_{\rm um}$
is the averaged value of 
$V^{(2)}_{{\bf k}, {\bf p}}$.
Using the definition of the order parameter in terms of the anomalous
operator averages,
Eq.~(\ref{suppl::Delta}),
we derive
\begin{eqnarray}
\langle \hat{H}^{(2)} \rangle 
\approx
- N_c \, \frac{\bar{V}_{\rm um}}{\bar{V}_1^2} \,
	\Delta_{\uparrow \xi} \,\Delta_{\downarrow \xi} + {\rm C.c.}
\end{eqnarray} 
This suggests that the self-consistent equations for
$s=(\uparrow, \xi)$
and
$s'=(\downarrow, \xi)$
become coupled. To account for this, we take the first of the two
equations~(\ref{suppl::Delta})
and add a term 
$V^{(2)}_\mathbf{p,k}
\langle \gamma^\dag_{\bar{\sigma}} \gamma_{\sigma} \rangle$
to its right-hand side
\begin{eqnarray} 
\Delta_{\mathbf{k} \sigma}
=
\frac{1}{N_c}
\sum_{\mathbf{p}}
	\!\left[
		V^{(1)}_\mathbf{p,k}
		\langle
			\gamma^\dag_{\mathbf{p}2\sigma}
			\gamma_{\mathbf{p}3\bar{\sigma}}
		\rangle
		\!+\!
  		V^{(2)}_{\mathbf{p,k}}
		\langle
			\gamma^\dag_{\mathbf{p}2\bar\sigma}
			\gamma_{\mathbf{p}3\sigma}
		\rangle
	\right],
\end{eqnarray} 
where we discarded the term with bands~1 and~4. Finally, using 
Eq.~(\ref{suppl::averages}),
we derive
\begin{eqnarray}
\label{suppl::eq::self_cons_gen1}
\Delta_\uparrow 
=
g \Delta_\uparrow 
\ln \left[\frac{E^*}{M(\mu, \Delta_\uparrow)} \right]
+
g_{\rm um} \Delta_\downarrow 
\ln \left[ \frac{E^*}{M(\mu, \Delta_\downarrow)} \right],
\\
\label{suppl::eq::self_cons_gen2}
\Delta_\downarrow 
=
g \Delta_\downarrow 
\ln \left[\frac{E^*}{M(\mu, \Delta_\downarrow)} \right]
+
g_{\rm um} \Delta_\uparrow 
\ln \left[ \frac{E^*}{M(\mu, \Delta_\uparrow)} \right].
\end{eqnarray}
To describe two remaining sectors,
$(\uparrow, \bar\xi)$
and
$(\downarrow, \bar\xi)$,
the identical set of equations should be used. In
Eqs.~(\ref{suppl::eq::self_cons_gen1})
and~(\ref{suppl::eq::self_cons_gen2}),
the quantity
$M(\mu, \Delta)$ 
effectively functions as the low-energy cutoff: if in a given sector
$\Delta > \mu$,
this sector remains undoped, and
$M(\mu, \Delta) = \Delta$;
when a sector accommodates finite doping
$\mu > \Delta$,
in such a situation
$M(\mu, \Delta) = \mu + \sqrt{\mu^2 - \Delta^2}$. 
Formally, this can be expressed as
\begin{eqnarray} 
M(\mu, \Delta) = (\mu+\sqrt{\mu^2-\Delta^2})\Theta(\mu - \Delta) +
\Delta\Theta(\Delta - \mu).
\end{eqnarray} 
Note also that in
Eqs.~(\ref{suppl::eq::self_cons_gen1})
and~(\ref{suppl::eq::self_cons_gen2})
we used the simplified notation
$\Delta_\uparrow \equiv \Delta_{\uparrow\xi}$
and
$\Delta_\downarrow \equiv \Delta_{\downarrow\xi}$.
The coupling constant is
$g_{\rm um} = \beta \bar{V}_{\rm um} \nu (\varepsilon_{\rm F})$,
where
$\nu (\varepsilon_{\rm F})$
is the density of states, and $\beta$ is a numerical coefficient of order
unity.

When the system is undoped, we can introduce
$\Delta_0$
as follows
$\mu = \Delta_\uparrow = \Delta_\downarrow \equiv \Delta_0$
[note that this is a redefinition of
$\Delta_0$
initially given by
Eq.~(\ref{suppl::eq::Delta0})].
In such a limit, both equations become identical
\begin{eqnarray}
\Delta_0
=
g (1 + \gamma) \Delta_0 \ln \left(\frac{E^*}{\Delta_0} \right),
\quad
\text{where}
\quad
\gamma = \frac{g_{\rm um}}{g}.
\end{eqnarray} 
This equation has one non-zero solution
\begin{eqnarray}
\label{suppl::eq::Delta0_umklapp}
\Delta_0 = E^* \exp\left( - \frac{1}{g(1+\gamma)} \right).
\end{eqnarray} 
We can see that the umklapp coupling increases 
$\Delta_0$.

\subsection{Doped state}

Now we discuss the doped system. Below we will consider two possibilities:
(i)~all four sectors are doped equally, and (ii)~three sectors remain
undoped, and all doping only enters a single sector. 
Let us start with (i). In
such a situation 
$\mu > \Delta_s = \Delta(x)$
for all four $s$.
Equations~(\ref{suppl::eq::self_cons_gen1})
and~(\ref{suppl::eq::self_cons_gen2})
become identical
\begin{eqnarray}
\Delta = g (1 + \gamma) \Delta
	\ln \left(\frac{E^*}{\mu + \sqrt{\mu^2 - \Delta^2}} \right),
\end{eqnarray} 
valid in all four sectors. The solution to this equation is similar to
Eq.~(\ref{suppl::DeltaDop}) 
\begin{eqnarray}
\Delta (x) =\Delta_0\sqrt{1-\frac{x}{x_0}},
\qquad
\mu=\Delta_0\left(1-\frac{x}{2x_0}\right),
\end{eqnarray} 
where we took into account that partial dopings equal to half of the total
doping:
$x_s = x/4$.
The expression for
$\mu (x)$
allows us to calculate 
$\Delta F(x)$
\begin{eqnarray}
\label{suppl::eq::free_en_metal}
\Delta F(x) = 4 \int_0^{x} \mu(x)\, dx
=
4 \Delta_0\,  x - \Delta_0 \frac{x^2}{ x_0}.
\end{eqnarray} 
This free energy is denoted as
$\Delta F_{\rm e}$
in the main text.

For case~(ii), the calculations are more complicated. We define
$\delta_\sigma (x)$
as follows
$\Delta_\sigma (x) = \Delta_0 [ 1 - \delta_\sigma (x)]$.
For definiteness, we assume that the sector 
$s = (\uparrow, \xi)$
is undoped, while
$s = (\downarrow, \xi)$
is doped. This means that
$\Delta_\uparrow > \mu > \Delta_\downarrow$. 
Two other sectors,
$(\uparrow, \bar\xi)$
and
$(\downarrow, \bar\xi)$,
are undoped, and decoupled from $s$ and $s'$. Therefore, they are
characterized by the order parameter 
$\Delta_0$,
given by 
Eq.~(\ref{suppl::eq::Delta0_umklapp})
\begin{eqnarray}
0 < \delta_\uparrow < m < \delta_\downarrow,
\quad
\text{where}
\quad
m = \frac{\Delta_0 - \mu}{\Delta_0}.
\end{eqnarray}
Let us introduce yet another quantity, 
$\delta S$,
as follows
\begin{eqnarray}
&&\mu + \sqrt{\mu^2 - \Delta_\downarrow^2}
=
\Delta_0 \left[
	1 - m + \sqrt{ (1-m)^2 - (1 - \delta_\downarrow)^2 }
	\right]
=
\Delta_0 ( 1 + \delta S ),
\\
&&\delta S = \sqrt{ (1-m)^2 - (1 - \delta_\downarrow)^2 } - m.
\end{eqnarray} 
The parameters
$\delta_\sigma$,
$\delta S$,
and $m$ are small in the limit of small doping $x$. However, they have
different degrees of smallness. Indeed, as we will see later
\begin{eqnarray}
\label{suppl::eq::O}
\delta_\sigma = O(m),
\quad
\delta S = O(m^{1/2}).
\end{eqnarray} 
These relations become important when we solve the self-consistent
equations in the limit of small doping.

Our goal is to solve the following equations
\begin{eqnarray}
(1 - \delta_\uparrow )
=
g (1 - \delta_\uparrow )
\left[
	\frac{1}{g(1+\gamma)}
	-
	\ln \left( 1 - \delta_\uparrow \right)
\right]
+
\gamma g (1 - \delta_\downarrow )
\left[
	\frac{1}{g(1+\gamma)}
	-
	\ln (1 + \delta S)
\right],
\\
(1 - \delta_\downarrow )
=
g (1 - \delta_\downarrow )
\left[
	\frac{1}{g(1+\gamma)}
	-
	\ln \left( 1 + \delta S \right)
\right]
+
\gamma g (1 - \delta_\uparrow )
\left[
	\frac{1}{g(1+\gamma)}
	-
	\ln (1 - \delta_\uparrow )
\right],
\end{eqnarray} 
to find 
$\delta_\sigma$
as a function of $m$, and then determine $m$ versus $x$. In the limit of
small $x$, we expand the self-consistent equations and, keeping in mind
Eq.~(\ref{suppl::eq::O}),
we derive
\begin{eqnarray}
(1 - \delta_\uparrow )
=
g (1 - \delta_\uparrow )
\left[ \frac{1}{g(1+\gamma)} + \delta_\uparrow \right]
+
\gamma g (1 - \delta_\downarrow )
\left[ \frac{1}{g(1+\gamma)} - \delta S + \frac{\delta S^2}{2} \right]
+ O(m^{3/2}),
\\
(1 - \delta_\downarrow )
=
g (1 - \delta_\downarrow )
\left[ \frac{1}{g(1+\gamma)} - \delta S + \frac{\delta S^2}{2} \right]
+
\gamma g (1 - \delta_\uparrow )
\left[
	\frac{1}{g(1+\gamma)}
	+
	\delta_\uparrow 
\right]
+ O(m^{3/2}).
\end{eqnarray} 
Simplifying, we obtain
\begin{eqnarray}
\delta_\uparrow
\approx
\left[ \frac{\delta_\uparrow}{1+\gamma} - g \delta_\uparrow \right]
+
\gamma \left[
	\frac{1}{1+\gamma} \delta_\downarrow 
	+
	g \left( \delta S - \frac{\delta S^2}{2} \right)
\right],
\\
\delta_\downarrow 
\approx
\left[
	\frac{\delta_\downarrow }{1+\gamma}
	+
	g \left(\delta S - \frac{\delta S^2}{2} \right)
\right]
+
\gamma \left[
	\frac{1}{1+\gamma}\delta_\uparrow 
	-
	g \delta_\uparrow 
\right],
\end{eqnarray} 
Next step:
\begin{eqnarray}
\left( \frac{\gamma}{1+\gamma} + g \right) \delta_\uparrow 
=
\frac{\gamma }{1+\gamma} \delta_\downarrow 
	+
	g \gamma \left( \delta S - \frac{\delta S^2}{2} \right),
\\
\left( \frac{\gamma }{1+\gamma} - g \gamma \right)\delta_\uparrow 
=
\frac{\gamma }{1+\gamma} \delta_\downarrow
-
g \left(\delta S - \frac{\delta S^2}{2} \right).
\end{eqnarray} 
Subtracting these two equations we derive
\begin{eqnarray}
g ( 1 + \gamma ) \delta_\uparrow
=
g ( 1 + \gamma ) \left(\delta S - \frac{\delta S^2}{2} \right)
\quad
\Leftrightarrow
\quad
\delta_\uparrow = \delta S - \frac{\delta S^2}{2}.
\end{eqnarray} 
Now
$\delta_\uparrow$
can be eliminated
\begin{eqnarray}
\left[ \frac{\gamma}{1+\gamma} + g (1 - \gamma) \right]
\left(\delta S - \frac{\delta S^2}{2} \right)
=
\frac{\gamma }{1+\gamma} \delta_\downarrow.
\end{eqnarray}
This relation is equivalent to
\begin{eqnarray} 
\delta S - \frac{\delta S^2}{2} = \alpha \delta_\downarrow,
\quad
\text{where}
\quad
\alpha = \left[ 1 + g (\gamma^{-1} - \gamma) \right]^{-1}. 
\end{eqnarray} 
Let us express
$\delta S$
in the limit of small doping
\begin{eqnarray}
&&\delta S = \sqrt{ (1-m)^2 - (1 - \delta_\downarrow)^2 } - m
=
\sqrt{(2 - m - \delta_\downarrow)(\delta_\downarrow - m)} - m
=
\sqrt{2 (\delta_\downarrow - m)} - m + O(m^{3/2}),
\\
&&\delta S^2 = 2 (\delta_\downarrow - m) + O(m^{3/2}).
\end{eqnarray} 
Therefore
\begin{eqnarray}
\delta S - \frac{\delta S^2}{2} 
=
\sqrt{2 (\delta_\downarrow - m)} - \delta_\downarrow + O(m^{3/2}).
\end{eqnarray} 
The self-consistent equation becomes
\begin{eqnarray}
\alpha \delta_\downarrow
=
\sqrt{2 (\delta_\downarrow - m)} - \delta_\downarrow + O(m^{3/2}).
\end{eqnarray} 
Its solution is
\begin{eqnarray}
\label{suppl::eq::delta_down}
\delta_\downarrow \approx m + \frac{(1+\alpha)^2}{2} m^2,
\quad
\delta_\uparrow = \alpha \delta_\downarrow
\approx
\alpha m + \frac{\alpha (1+\alpha)^2}{2} m^2.
\end{eqnarray} 
Let us check the consistency of these relations with known results in the
$\alpha = 0$
limit. In this case
\begin{eqnarray}
\label{suppl::eq::alpha0}
\delta_\downarrow \approx m + \frac{1}{2} m^2,
\quad
\delta_\uparrow = 0.
\end{eqnarray}
At the same time, 
Eqs.~(\ref{suppl::DeltaDop})
in the regime of small $x$ can be written as
\begin{eqnarray}
m = \frac{x}{2x_0},
\quad
\delta_\downarrow (x) = \frac{\Delta_0 - \Delta (x)}{\Delta_0}
\approx
\frac{x}{2x_0} + \frac{x^2}{8x_0^2}.
\end{eqnarray} 
We can now exclude $x$ to obtain
\begin{eqnarray}
\delta_\downarrow (x) \approx m + \frac{m^2}{2},
\end{eqnarray} 
which coincides with
Eq.~(\ref{suppl::eq::alpha0}).

The final step is to add doping into the formalism. To this end, we write
\begin{eqnarray}
4x = 2 \nu_F \int_{\Delta_\downarrow}^{\mu} d \varepsilon 
	\frac{\varepsilon}{\sqrt{\varepsilon^2 - \Delta_\downarrow^2}},
\end{eqnarray} 
where $4x$ is the doping per unit cell, 
$\nu_F = t_0/(\sqrt{3}\pi t^2)$
is the density of states per unit cell for each single Fermi surface sheet
(there are four Fermi surface sheets),
\begin{eqnarray}
x = \frac{\nu_F}{2} \sqrt{\mu^2 - \Delta_\downarrow^2}
= \frac{\nu_F \Delta_0}{2} \sqrt{(1-m)^2 - (1-\delta_\downarrow)^2}
= \frac{x_0}{2} \sqrt{(1-m)^2 - (1-\delta_\downarrow)^2},
\end{eqnarray}
where
$x_0 = \nu_F \Delta_0$.
It is possible to show that
\begin{eqnarray}
\label{suppl::eq::x2_vs_mu}
4x^2 = x_0^2 (\delta_\downarrow - m)(2 - m - \delta_\downarrow)
\quad
\Rightarrow
\quad
4 x^2 = (1+\alpha)^2 x_0^2 m^2 + O(m^3).
\end{eqnarray} 
Deriving the latter relation we used
Eq.~(\ref{suppl::eq::delta_down}),
which, among other things, demonstrates that
$\delta_\downarrow - m = O(m^2)$.
Equation~(\ref{suppl::eq::x2_vs_mu})
allows us to establish the following connection between doping and the
chemical potential
\begin{eqnarray}
m = \frac{2x}{(1+\alpha) x_0} + O(x^2)
\quad
\Leftrightarrow
\quad
\mu = \Delta_0 \left( 1 - \frac{2x}{(1+\alpha)x_0} \right) + O(x^2).
\end{eqnarray} 
\begin{figure}[t!]
\includegraphics[width=9cm]{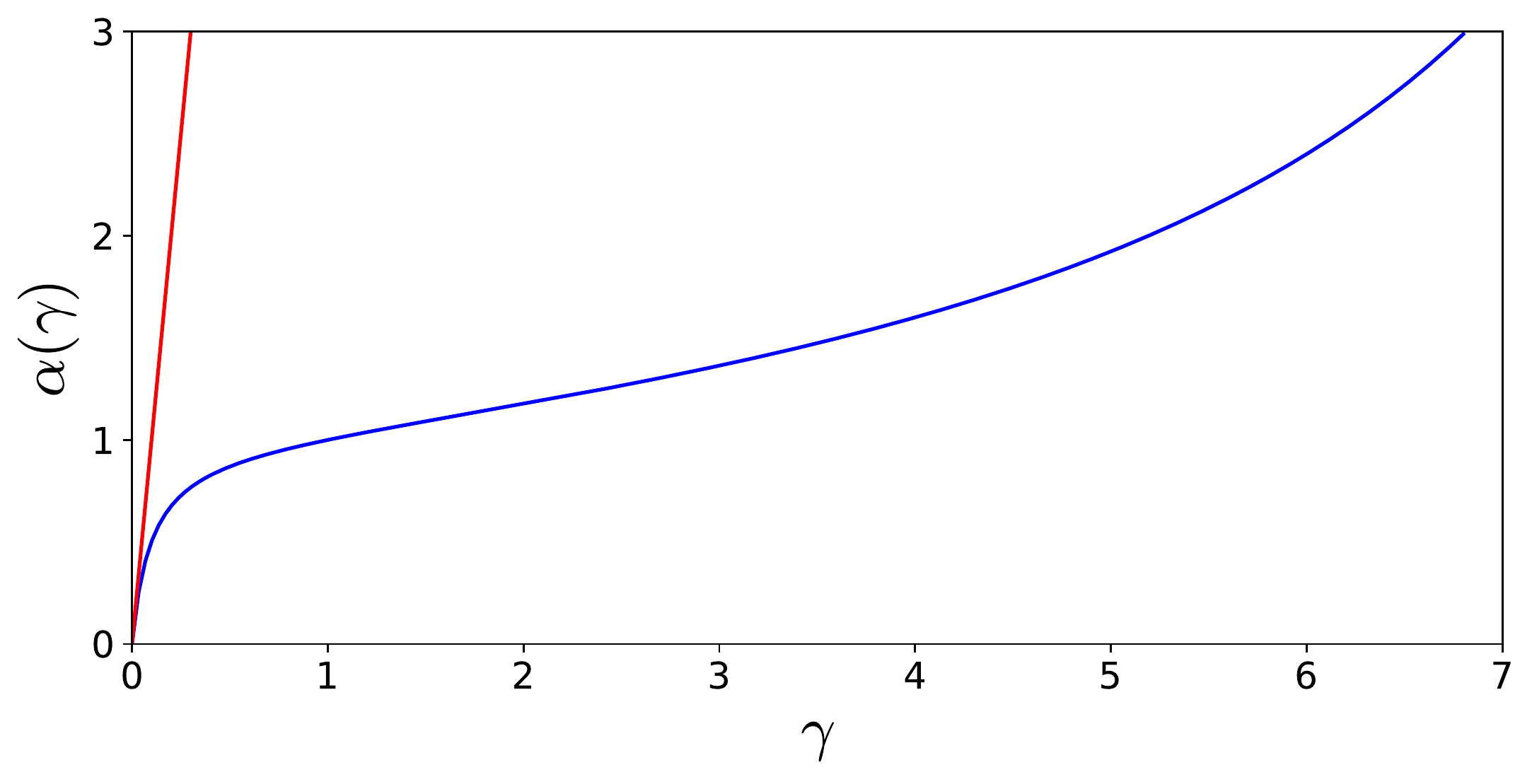}
\caption{The function
$\alpha (\gamma )$
for 
$g=0.1$
is shown by the blue curve. The straight (red) line is
$\gamma / g$.
\label{suppl::fig::alpha}
}
\end{figure}
Integrating
$\mu(x)$,
we obtain
\begin{eqnarray}
\label{suppl::eq::free_en_qm}
\Delta F_{\rm qm}
=
4 \Delta_0 x - \left( \frac{4 \Delta_0}{1+\alpha}\right) \frac{x^2}{x_0}.
\end{eqnarray} 
In the limit
$\alpha \rightarrow 0$
we recover the expression for
$\Delta F_{\rm qm}$
given in the main text [see after Eq.~(30)]. The free
energy~(\ref{suppl::eq::free_en_qm})
must be compared against the free energy given by
Eq.~(\ref{suppl::eq::free_en_metal}).
We see that the quarter-metal is stable if 
$(1+\alpha)^{-1} > 1/4$.
Equivalently, 
\begin{eqnarray}
\label{suppl::eq::qm_stable}
\text{quarter-metal is stable when}
\quad
\alpha (\gamma) < 3.
\end{eqnarray} 
To understand what the latter requirement entails, let us examine
Fig.~\ref{suppl::fig::alpha},
which shows 
$\alpha (\gamma)$
for
$g=0.1$.
We see that 
$\alpha < 3$
as long as
$\gamma = g_{\rm um}/g < 6.8$.
That is, for 
$g = 0.1$,
the umklapp satisfying
\begin{eqnarray}
\label{suppl::eq::example}
g_{\rm um} < 0.68,
\end{eqnarray} 
does not violate the stability of the quarter-metal.

We note that
Eq.~(\ref{suppl::eq::qm_stable})
is not the absolute stability criterion, rather it describes the stability
of the quarter-metal against the transition into an ordinary metal, when
all four sectors are doped equally. A comprehensive investigation of the
stability goes well beyond the present study, and, most likely, requires
input from experiments.

It is interesting to note that perturbation theory in powers of small
$\gamma$ strongly underestimates the stability range of the quarter-metal.
To demonstrate this, we expand the
expression~(\ref{suppl::eq::free_en_qm})
for
$\Delta F_{\rm qm}$
in powers of $\alpha$
\begin{eqnarray}
\Delta F_{\rm qm} \approx 4\Delta_0 x - 4 \Delta_0 \frac{x^2}{x_0} 
+ 4 \Delta_0 \frac{\alpha x^2}{x_0}.
\end{eqnarray} 
Since at 
$\gamma \rightarrow 0$,
the following holds
$\alpha \approx \gamma/g = g_{\rm um}/g^2$,
the expression for
$\Delta F_{\rm qm}$
can be approximated as
\begin{eqnarray}
\Delta F_{\rm qm} \approx
4 \Delta_0 x - 4 \Delta_0 \frac{x^2}{ x_0} 
	\left( 1 - \frac{g_{\rm um}}{g^2} \right),
\end{eqnarray} 
If we use this expression, instead of the more accurate
Eq.~(\ref{suppl::eq::free_en_qm}),
we could (erroneously) conclude that the quarter-metal is stable when
$( 1 - {g_{\rm um}}/{g^2}) > 1/4$.
This inequality can be transformed to
\begin{eqnarray} 
\label{suppl::eq::pert_theor_stability}
\frac{\gamma}{g} = \frac{g_{\rm um}}{g^2} < \frac{3}{4}
\quad
\Leftrightarrow
\quad
g_{\rm um} < \frac{3 g^2}{4}.
\end{eqnarray} 
In
Fig.~\ref{suppl::fig::alpha}
we can see the low-$\gamma$ approximation
$\alpha (\gamma ) \approx \gamma/g$
as a (red) straight line. We see that, at low $g$, this approximation works
only at very small $\gamma$; while for larger
$\gamma$ (larger
$g_{\rm um}$)
it is completely useless. Thus, we conclude that the replacement
$[1 + \alpha (\gamma)]^{-1} \rightarrow ( 1 - \gamma/g )$
artificially shrinks the stability range of the quarter-metal. Indeed, the
requirement~(\ref{suppl::eq::pert_theor_stability})
is very strict: at 
$g=0.1$,
as in
Fig.~\ref{suppl::fig::alpha},
Eq.~(\ref{suppl::eq::pert_theor_stability})
demand that
$g_{\rm um} < 0.0075$,
cf.
Eq.~(\ref{suppl::eq::example}).
This is the origin of the serious disparity between the stability condition
derived in the main text using simple perturbation theory and more the
sophisticated
criterion~(\ref{suppl::eq::qm_stable}).

\end{widetext}

\end{document}